\documentclass{aastex63}

\received{}
\revised{}
\accepted{}
\submitjournal{ApJ}

\shorttitle{Carbon Chain Chemistry in Hot-Core Regions around MYSOs}
\shortauthors{Taniguchi et al.}
\begin{document}

\title{Carbon Chain Chemistry in Hot-Core Regions around Three Massive Young Stellar Objects Associated with 6.7 GHz Methanol Masers}

\correspondingauthor{Kotomi Taniguchi}
\email{kotomi.taniguchi@gakushuin.ac.jp}

\author[0000-0003-4402-6475]{Kotomi Taniguchi}
\affiliation{Department of Physics, Faculty of Science, Gakushuin University, Mejiro, Toshima, Tokyo 171-8588, Japan}

\author[0000-0002-4649-2536]{Eric Herbst}
\affiliation{Department of Astronomy, University of Virginia, Charlottesville, VA 22904, USA}
\affiliation{Department of Chemistry, University of Virginia, Charlottesville, VA 22903, USA}

\author[0000-0001-7031-8039]{Liton Majumdar}
\affiliation{School of Earth and Planetary Sciences, National Institute of Science Education and Research, HBNI, Jatni 752050, Odisha, India}

\author[0000-0003-1481-7911]{Paola Caselli}
\affiliation{Max-Planck-Institute for Extraterrestrial Physics (MPE), Giessenbachstr, 1, D-85748 Garching, Germany}

\author[0000-0002-3389-9142]{Jonathan C. Tan}
\affiliation{Department of Astronomy, University of Virginia, Charlottesville, VA 22904, USA}
\affiliation{Department of Space, Earth \& Environment, Chalmers University of Technology, 412 93  Gothenburg, Sweden}

\author{Zhi-Yun Li}
\affiliation{Department of Astronomy, University of Virginia, Charlottesville, VA 22904, USA}

\author[0000-0002-1054-3004]{Tomomi Shimoikura}
\affiliation{Faculty of Social Information Studies, Otsuma Women's University, Sanban-cho, Chiyoda, Tokyo 102-8357, Japan}

\author[0000-0001-8058-8577]{Kazuhito Dobashi}
\affiliation{Department of Astronomy and Earth Sciences, Tokyo Gakugei University, Nukuikitamachi, Koganei, Tokyo 184-8501, Japan}

\author[0000-0001-5431-2294]{Fumitaka Nakamura}
\affiliation{National Astronomical Observatory of Japan (NAOJ), National Institutes of Natural Sciences, Osawa, Mitaka, Tokyo 181-8588, Japan}
\affiliation{Department of Astronomical Science, School of Physical Science, SOKENDAI (The Graduate University for Advanced Studies), Osawa, Mitaka, Tokyo 181-8588, Japan}
\affiliation{The University of Tokyo, Hongo, Bunkyo, Tokyo 113-0033, Japan}

\author[0000-0003-0769-8627]{Masao Saito}
\affiliation{National Astronomical Observatory of Japan (NAOJ), National Institutes of Natural Sciences, Osawa, Mitaka, Tokyo 181-8588, Japan}
\affiliation{Department of Astronomical Science, School of Physical Science, SOKENDAI (The Graduate University for Advanced Studies), Osawa, Mitaka, Tokyo 181-8588, Japan}

%% Note that the \and command from previous versions of AASTeX is now
%% depreciated in this version as it is no longer necessary. AASTeX 
%% automatically takes care of all commas and "and"s between authors names.

%% AASTeX 6.3 has the new \collaboration and \nocollaboration commands to
%% provide the collaboration status of a group of authors. These commands 
%% can be used either before or after the list of corresponding authors. The
%% argument for \collaboration is the collaboration identifier. Authors are
%% encouraged to surround collaboration identifiers with ()s. The 
%% \nocollaboration command takes no argument and exists to indicate that
%% the nearby authors are not part of surrounding collaborations.

%% Mark off the abstract in the ``abstract'' environment. 
\begin{abstract}
We have carried out observations of CCH ($N=1-0$), CH$_{3}$CN ($J=5-4$), and three $^{13}$C isotopologues of HC$_{3}$N ($J=10-9$) toward three massive young stellar objects (MYSOs), G12.89+0.49, G16.86--2.16, and G28.28--0.36, with the Nobeyama 45-m radio telescope.
Combined with previous results on HC$_{5}$N, the column density ratios of $N$(CCH)/$N$(HC$_{5}$N), hereafter the CCH/HC$_{5}$N ratios, in the MYSOs are derived to be $\sim 15$. 
This value is lower than that in a low-mass warm carbon chain chemistry (WCCC) source by more than one order of magnitude.
We compare the observed CCH/HC$_{5}$N ratios with hot-core model calculations \citep{2019ApJ...881...57T}.
The observed ratios in the MYSOs can be best reproduced by models when the gas temperature is $\sim 85$ K, which is higher than in L1527, a low-mass WCCC source ($\sim 35$ K). 
These results suggest that carbon-chain molecules detected around the MYSOs exist at least partially in higher temperature regions than those in low-mass WCCC sources.
There is no significant difference in column density among the three $^{13}$C isotopologues of HC$_{3}$N in G12.89+0.49 and G16.86-2.16, while HCC$^{13}$CN is more abundant than the others in G28.28--0.36.
We discuss carbon-chain chemistry around the three MYSOs based on the CCH/HC$_{5}$N ratio and the $^{13}$C isotopic fractionation of HC$_{3}$N.
\end{abstract}

%% Keywords should appear after the \end{abstract} command. 
%% See the online documentation for the full list of available subject
%% keywords and the rules for their use.

%% From the front matter, we move on to the body of the paper.
%% Sections are demarcated by \section and \subsection, respectively.
%% Observe the use of the LaTeX \label
%% command after the \subsection to give a symbolic KEY to the
%% subsection for cross-referencing in a \ref command.
%% You can use LaTeX's \ref and \label commands to keep track of
%% cross-references to sections, equations, tables, and figures.
%% That way, if you change the order of any elements, LaTeX will
%% automatically renumber them.
%%
%% We recommend that authors also use the natbib \citep
%% and \citet commands to identify citations.  The citations are
%% tied to the reference list via symbolic KEYs. The KEY corresponds
%% to the KEY in the \bibitem in the reference list below. 

\section{Introduction} \label{sec:intro}

Approximately 200 molecules have been identified in the interstellar medium and circumstellar shells \citep{2018ApJS..239...17M}.
Around 40\% of the interstellar molecules are accounted for by unsaturated carbon-chain molecules.
Carbon-chain molecules, such as CCS and HC$_{5}$N, are generally known to be abundant in cold ($T \approx 10$ K) starless/prestellar cores \citep[e.g.,][]{1992ApJ...392..551S, 1998ApJ...506..743B}.
They are formed in the gas phase via ion-molecule reactions and exothermic neutral-neutral reactions without reaction energy barriers. 
Their abundances decrease in star-forming cores \citep{1992ApJ...392..551S, 1998ApJ...506..743B}, due to their destruction and/or depletion onto dust grains in cold high density regions just before protostars form.
Based on the above characteristics, these carbon-chain species are also called ``early-type species".

Around protostars, partially saturated complex organic molecules (COMs), which consist of more than six atoms and are rich in hydrogen atoms, are abundant \citep{2009ARA&A..47..427H}.
These regions are referred to as hot cores and hot corinos in high-mass and low-mass star-forming regions, respectively.
The COMs are formed on dust grains in the cold starless core and warm-up phases \citep{2006A&A...457..927G, 2008ApJ...682..283G}.
Gas-phase reactions also contribute to the formation of COMs in both warm regions \citep[e.g.,][]{2019MNRAS.482.3567S} and in cold environments \citep{2015MNRAS.449L..16B}.
Although general formation processes of COMs have been proposed, recent ALMA observations have revealed complexities of COM chemistry around protostars both in high-mass and low-mass star-forming regions \citep[][and therein]{2020arXiv200607071J}.
For example, a chemical differentiation between N-bearing COMs and O-bearing COMs has been found \citep[e.g.,][]{2015ApJ...806..239C}.
Even among O-bearing COMs, there is a spatial segregation between COMs containing a C-O-C structure and those containing a C-OH bond \citep{2018A&A...620L...6T}.

In contrast to hot corino chemistry, warm carbon chain chemistry (WCCC) was proposed initially around some low-mass protostars, such as L1527 and IRAS 15398--3359 \citep{2013ChRv..113.8981S}.
In WCCC sources, carbon-chain molecules are efficiently formed from methane (CH$_{4}$) which is desorbed from dust grains at temperatures of $\sim 25-30$ K \citep{2008ApJ...681.1385H}.
Reactions between CH$_{4}$ molecules and C$^{+}$ ions initiate carbon-chain formation in the lukewarm gas ($T \approx 25-30$ K).
In high-mass star-forming regions, observational studies concerning carbon-chain chemistry are less well developed.
It is unknown whether carbon-chain molecules are formed anew around massive young stellar objects (MYSOs), and if they are formed, whether these carbon-chain molecules exist in higher temperature regions than those in low-mass WCCC sources.

\citet{2014MNRAS.443.2252G} carried out survey observations of HC$_{5}$N toward MYSOs containing 6.7 GHz CH$_{3}$OH masers.
They detected HC$_{5}$N in 35 sources out of 79 MYSOs using the $J=12-11$ line (31.951777 GHz; $E_{\rm {up}}/k = 9.97$ K) with the 34-m radio telescope at Tidbinbilla.
However, the detected HC$_{5}$N lines seem to come mainly from cold envelopes ($T \approx 10$ K), not from inner hot regions, because of the large beam size (0.95\arcmin) and the low energy level of the observed line.
\citet{2018ApJ...854..133T} detected HC$_{5}$N in 14 high-mass protostellar objects (HMPOs) out of 35 sources using the $J = 16 - 15$ line (42.60215 GHz; $E_{\rm {up}}/k = 17.4$ K) with the Nobeyama 45-m telescope and a beam size of 37\arcsec.
However, this line also may come from cold envelopes, and it was unclear whether HC$_{5}$N is in warm ($T \approx 25$ K) and/or hot ($T \gtrsim 100$ K) regions.

\citet{2017ApJ...844...68T} also carried out observations of HC$_{5}$N toward four MYSOs with the Green Bank 100-m and the Nobeyama 45-m radio telescopes.
They detected several lines with high upper state energies that are difficult to be excited in cold regions (e.g., $J=39-38$; $E_{\rm {up}}/k = 99.7$ K).
These results suggest that HC$_{5}$N exists in warm and/or hot regions around MYSOs.
However, their observations were conducted with single-dish telescopes and it could not be strongly concluded that the detected HC$_{5}$N around MYSOs exists in higher temperature regions than found in low-mass WCCC sources.
\citet{2018ApJ...866...32T} investigated the spatial distributions of HC$_{3}$N and HC$_{5}$N in the G28.28--0.36 MYSO, and their spatial distributions are positively correlated with the dust continuum emission.
These results support the possibility of gas-phase cyanopolyyne formation (HC$_{2n+1}$N, $n=1,2,3,...$) in warm regions from dust-origin precursors.

Recently, \citet{2019ApJ...881...57T} conducted gas-grain chemical network simulations in order to investigate the carbon-chain chemistry around MYSOs in detail.
Based on these time-dependent calculations, they suggested that there are largely two types of carbon-chain species.
During the warm-up period, cyanopolyynes, which are relatively stable species, are formed in the gas phase, depleted onto dust grains and condensed in ice mantles during the lukewarm phase (25 K $<T<100$ K).
After their sublimation temperature is reached ($T\geq100$ K), cyanopolyynes are desorbed from dust grains and their gas-phase abundances show peak values.
On the other hand, CCH is efficiently formed and shows its peak abundance in a WCCC-like region just after the temperature reaches 25 K, at which temperature the CH$_{4}$ molecules are desorbed from dust grains. 
The CCH radicals are rapidly formed by the electron recombination reaction of C$_{2}$H$_{3}^{+}$, which is formed by the reaction between CH$_{4}$ and C$^{+}$.
After the temperature exceeds 70 K, CCH is efficiently destroyed by atomic oxygen (O). 
The main destroyer of CCH becomes H$_{2}$ at temperatures above 90 K, and destruction becomes much faster.
Motivated by this model calculation \citep{2019ApJ...881...57T}, we infer that the CCH/HC$_{5}$N ratio can be used as a temperature probe over the range where these carbon-chain molecules exist.
This ratio is expected to decrease with an increase in temperature.
Hence, if HC$_{5}$N around MYSOs exists to a greater extent in higher temperature regions than in lukewarm regions around low-mass WCCC sources, the CCH/HC$_{5}$N ratios in MYSOs will be lower than those in WCCC sources.

In this paper, we report new observations of CCH ($N=1-0$), CH$_{3}$CN ($J=5-4$), and three $^{13}$C isotopologues of HC$_{3}$N ($J=10-9$) toward three MYSOs (G12.89+0.49, G16.86--2.16, and G28.28--0.36) with the Nobeyama 45-m radio telescope.
We describe the observations in Section \ref{sec:obs}.
The spectra of the above three species are presented in Section \ref{sec:res}, and analytical methods and results are summarized in Section \ref{sec:ana}.
In combination with previous results on HC$_{5}$N reported by \citet{2017ApJ...844...68T}, we derive the CCH/HC$_{5}$N ratios toward the three MYSOs.
These CCH/HC$_{5}$N ratios toward the three MYSOs are compared with those in a low-mass WCCC source L1527 in Section \ref{sec:dis1}, and with the model calculation in Section \ref{sec:dis2}.
In Section \ref{sec:dis3}, the differences in column density among the three $^{13}$C isotopologues of HC$_{3}$N, namely the $^{13}$C isotopic fractionation, and possible main formation pathways of HC$_{3}$N are presented. 
We discuss carbon-chain chemistry around the three MYSOs based on the CCH/HC$_{5}$N ratio and the $^{13}$C isotopic fractionation of HC$_{3}$N in Section \ref{sec:dis4}.
The main conclusions of this paper are summarized in Section \ref{sec:con}.

\section{Observations} \label{sec:obs}

\begin{deluxetable*}{lccccccc}
\tablenum{1}
\tablecaption{Summary of target sources \label{tab:source}}
\tablewidth{0pt}
\tablehead{
		\colhead{Source} & \colhead{R.A.} & \colhead{Decl.}  & \colhead{$D$} &  \colhead{$V_{\rm {sys}}$\tablenotemark{d}} & \colhead{$L$} & \colhead{$M_{\rm {clump}}$} & \colhead{Spectral}\\
		 \colhead{}           & \colhead{(J2000)} & \colhead{(J2000)} &  \colhead{(kpc)} & \colhead{(\,km\,s$^{-1}$)} & \colhead{(L$_{\sun}$)} & \colhead{(M$_{\sun}$)} & \colhead{Type}
}
\startdata
		G12.89+0.49 & $18^{\rm {h}}11^{\rm {m}}$51\fs4 & -17\degr31\arcmin30\arcsec &  2.94\tablenotemark{a} & 33.3 & $3.3 \times 10^{4}$\tablenotemark{e} & 1505\tablenotemark{e} & B0\tablenotemark{g} \\ 
		G16.86--2.16  & $18^{\rm {h}}29^{\rm {m}}$24\fs4 & -15\degr16\arcmin04\arcsec & 1.7\tablenotemark{b} & 17.8 & $3.2 \times 10^{3}$\tablenotemark{f} & 437\tablenotemark{b} & B2\tablenotemark{g} \\
		G28.28--0.36  & $18^{\rm {h}}44^{\rm {m}}$13\fs3 & -04\degr18\arcmin03\arcsec & 3.0\tablenotemark{c} & 48.9  & ...\tablenotemark{h} & $723^{+124}_{-116}$\tablenotemark{h} & ... \\
\enddata
		\tablenotetext{a}{Taken from \citet{2013AA...553A.117I}.}
		\tablenotetext{b}{Taken from \citet{2015MNRAS.446.3461U}.}
		\tablenotetext{c}{Taken from \citet{2014MNRAS.443.2252G}.}
		\tablenotetext{d}{Taken from \citet{2006MNRAS.367..553P}.}
		\tablenotetext{e}{The mass and luminosity at the distance of 3.3 kpc are taken from \citet{2014ApJ...790...84L}. We converted these values into those at a distance of 2.94 kpc. The luminosity is indicated as $L_{\rm {IR}}$ in \citet{2014ApJ...790...84L}.}
		\tablenotetext{f}{The luminosity of G16.86--2.16 at a distance of 1.9 kpc is taken from \citet{2006MNRAS.367..553P}. We converted it into the value at a distance of 1.7 kpc.}
		\tablenotetext{g}{The approximate spectral types for a given luminosity are given in Table 1 in \citet{1973AJ.....78..929P}.}
		\tablenotetext{h}{The mass at a distance of 3.29 kpc is taken from \citet{2011ApJ...743...56C}. We converted it into that at a distance of 3.0 kpc. The bolometric luminosity could not be derived, because a clear counterpart is not present or because confusion/blending precludes measuring a reliable flux for the EGO counterpart. The dust temperature was assumed to be 28 K.}
\end{deluxetable*}

The observations presented in this paper were carried out with the Nobeyama 45-m radio telescope in 2019 April and May (2018--2019 semester\footnote{Proposal ID; SP189002, PI; Kotomi Taniguchi}).
Our three target MYSOs are the sources observed by \citet{2017ApJ...844...68T, 2018ApJ...866..150T}.
The properties of the target sources are summarized in Table \ref{tab:source}.
These sources were selected with the following criteria:
\begin{enumerate}
\item The source declination is above $-21\degr$,
\item the distance ($D$) is within 3 kpc, and
\item CH$_{3}$CN was detected with the Mopra telescope \citep[$\int T_{\rm{mb}}dv > 0.5$ K km\,s$^{-1}$ for the $J_{K}=5_{0}-4_{0}$ line;][]{2006MNRAS.367..553P}.
\end{enumerate}
In addition, the 6.7 GHz CH$_{3}$OH maser and molecular outflows are associated with all of the sources \citep{2008AJ....136.2391C, 2016AJ....152...92L}.
The 6.7 GHz CH$_{3}$OH maser is known as a sign of massive star formation \citep{2015MNRAS.446.3461U}.
The G28.28--0.36 MYSO is categorized as a GLIMPSE Extended Green Object \citep[EGO;][]{2011ApJ...743...56C}.

We used the FOREST receiver \citep{2016SPIE.9914E..1ZM} in the on-on observing mode\footnote{\url{https://www.nro.nao.ac.jp/~nro45mrt/html/obs/nobs/scan.html\#on-on}}.
The integration time for each scan was 20 seconds.
The off-source positions for calibrating observations were set at $+15\arcmin$ away in declination.
The main beam efficiency ($\eta_{\rm {mb}}$) and the beam size (HPBW) at 86 GHz of this receiver are $\sim50$\% and 18\arcsec, respectively.
The system temperatures were between $\sim 200$ K and $\sim 250$ K, depending on weather conditions and elevation.

We used the SAM45 FX-type digital correlator in a frequency setup, the bandwidth and frequency resolution of which are 500 MHz and 122.07 kHz, respectively.
This frequency resolution corresponds to a velocity resolution of 0.4\,km\,s$^{-1}$ at the observed frequency range.
We conducted 2-channel binning in the final spectra, and the velocity resolution of the final spectra is $\sim 0.8$\,km\,s$^{-1}$.

The telescope pointing was checked every 1.5 hr by observing the SiO maser line ($J=1-0$; 43.12203 GHz) from OH39.7+1.5 at ($\alpha_{2000}$, $\delta_{2000}$) = ($18^{\rm {h}}56^{\rm {m}}$03\fs88, +06\degr38\arcmin49\farcs8).
The H40 receiver was used for the pointing observations.
The pointing errors were within $3\arcsec$.

\section{Results and Analyses} \label{sec:RandA}

\subsection{Results} \label{sec:res}

We analyzed spectra with the Java NEWSTAR, which is a software package for data reduction and analyses of the Nobeyama 45-m radio telescope\footnote{\url{https://www.nro.nao.ac.jp/~jnewstar/html/}}.
The total integration times are approximately 9 hours, 6 hours, and 7 hours, for G12.89+0.49, G16.86--2.16, and G28.28--0.36, respectively.

\begin{figure*}[!th]
\figurenum{1}
 \begin{center}
  \includegraphics[bb=0 40 788 818, scale=0.5]{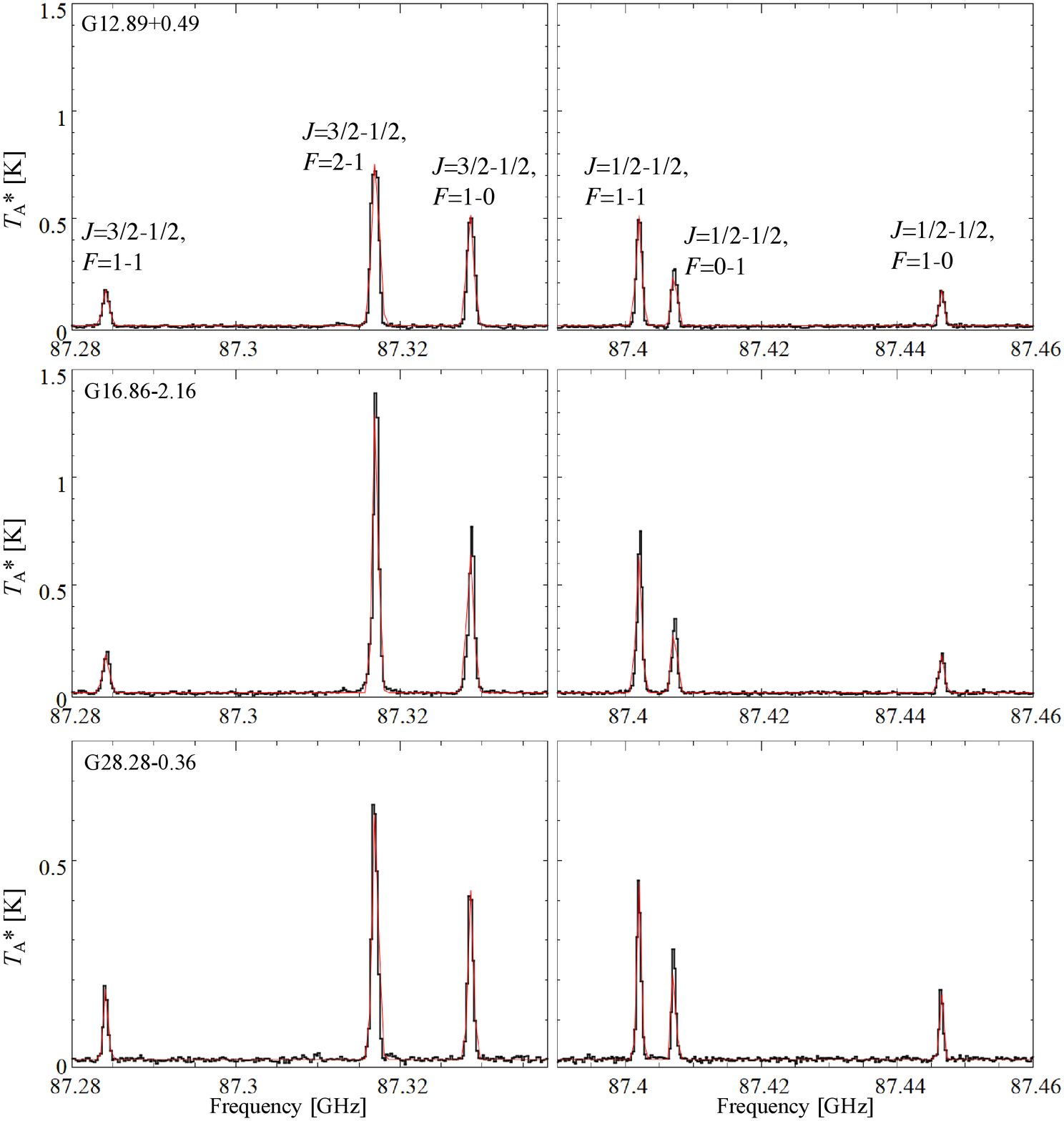}
 \end{center}
\caption{Spectra of the CCH ($N=1-0$) lines toward the three MYSOs obtained with the Nobeyama 45-m radio telescope. The red curves indicate the results of the Gaussian fit. $J= N \pm \frac{1}{2}$ (except for $N=0$, when $J=\frac{1}{2}$ only), and $F=J \pm \frac{1}{2}$.  $J$ couples electronic spin with rigid rotation, while $F$ couples nuclear spin.\label{fig:CCH}}
\end{figure*}

\begin{figure*}[!th]
\figurenum{2}
 \begin{center}
  \includegraphics[bb=0 40 685 652, scale=0.5]{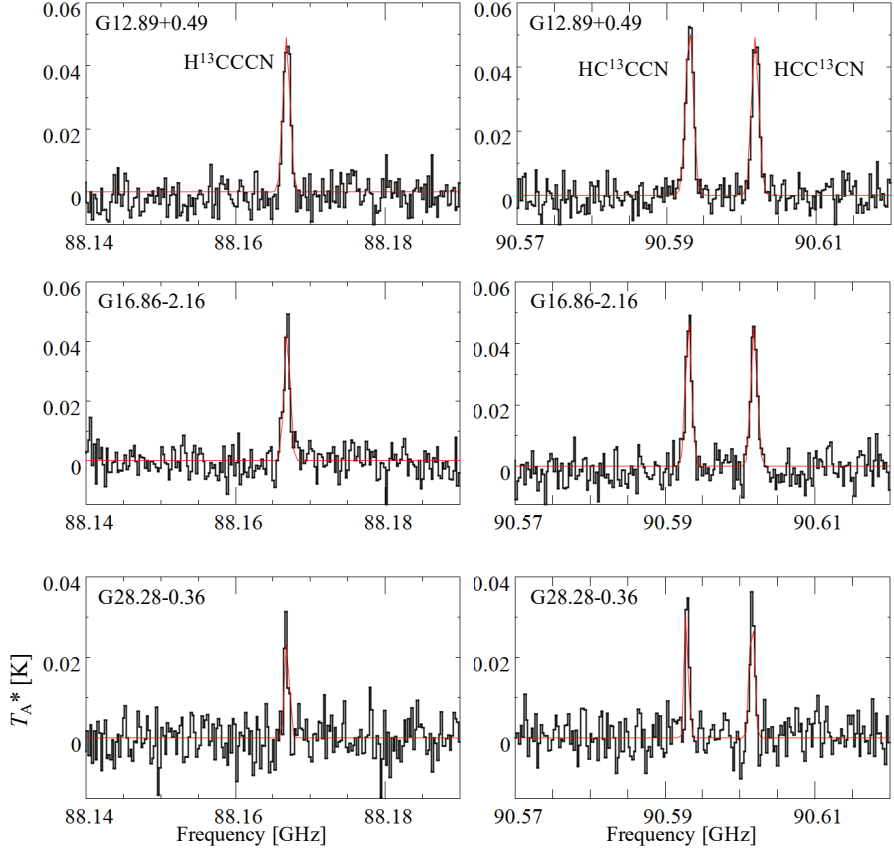}
 \end{center}
\caption{Spectra of three $^{13}$C isotopologues of HC$_{3}$N ($J=10-9$) toward the three MYSOs obtained with the Nobeyama 45-m radio telescope. The red curves indicate the results of the Gaussian fit. \label{fig:HC3N}}
\end{figure*}

\begin{figure*}[!th]
\figurenum{3}
 \begin{center}
  \includegraphics[bb=0 40 974 517, scale=0.5]{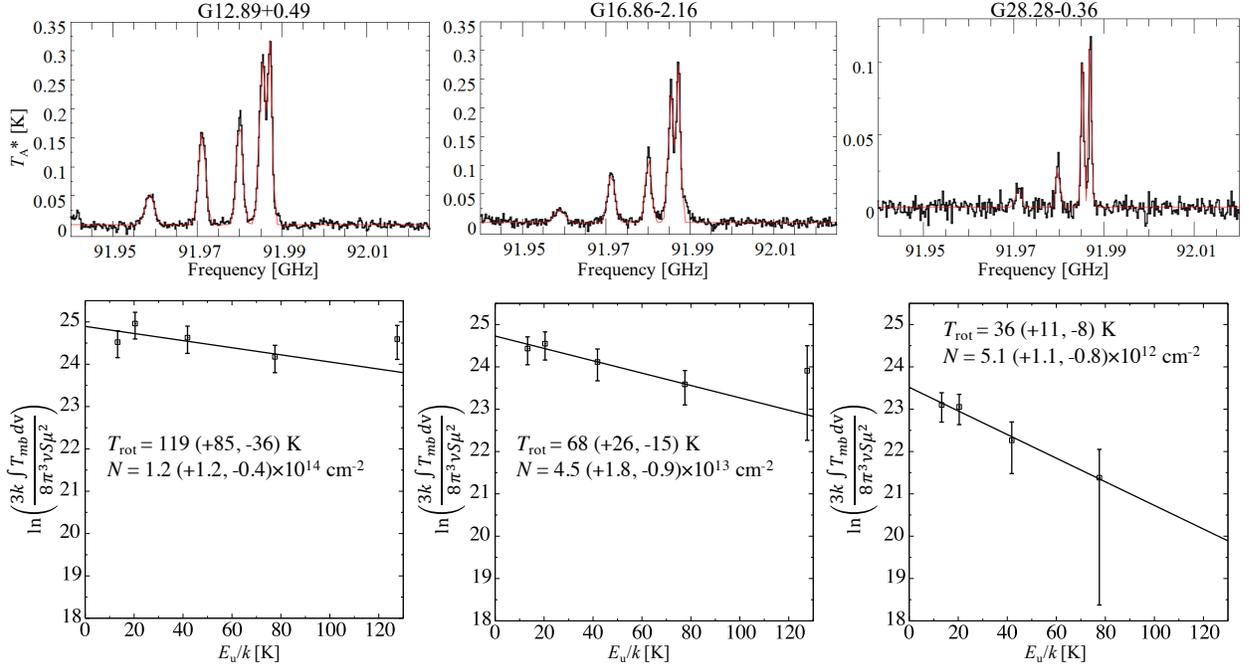}
 \end{center}
\caption{The top panels show spectra of CH$_{3}$CN ($J=5-4$, $K$-ladder lines) toward the three MYSOs obtained with the Nobeyama 45-m radio telescope. The red curves show the results of the Gaussian fitting of spectra. The bottom panels show the rotational diagram for each source. The error bars for each data point represent $3\sigma$ errors. \label{fig:CH3CN}}
\end{figure*}

Figures \ref{fig:CCH}--\ref{fig:CH3CN} show spectra of CCH ($N=1-0$), three $^{13}$C isotopologues of HC$_{3}$N ($J=10-9$), and CH$_{3}$CN ($J=5-4$), respectively, toward the three MYSOs.
The rms noise levels are 4--5 mK in $T_{\rm {A}}^{\ast}$ scale in all of the sources.
We fitted the spectra with a Gaussian profile.
The obtained spectral line parameters are summarized in Table \ref{tab:line}.
We applied a main beam efficiency ($\eta_{\rm {mb}}$) of 50\%, 46\%, and 48\% to the CCH, CH$_{3}$CN, and three $^{13}$C isotopologues of HC$_{3}$N lines, respectively, for calculation of $T_{\rm {mb}}$ values.
These $\eta_{\rm {mb}}$ values were calculated by extrapolation using the reported values at 86 GHz and 110 GHz\footnote{\url{https://www.nro.nao.ac.jp/~nro45mrt/html/prop/eff/eff2019.html}}.

The observed radial velocities ($V_{\rm {LSR}}$) of each line are consistent with the systemic velocities of each source (Table \ref{tab:source}) within the errors.
The $J_{K}=5_{3}-4_{3}$ line of CH$_{3}$CN in G28.28--0.36 shows a slightly larger $V_{\rm {LSR}}$ value compared with the systemic velocity, but this is due to the low signal-to-noise (S/N) ratio of this line.
 
Six lines of the hyperfine and fine structures of CCH have been detected from the three MYSOs as shown in Figure \ref{fig:CCH}.
Their line widths ($\Delta v$) obtained by the Gaussian fitting are $\sim 3-3.5$\,km\,s$^{-1}$ in G12.89+0.49 and G16.86--2.16, and $\sim 2.5-3$\,km\,s$^{-1}$ in G28.28--0.36, respectively (see Table \ref{tab:line}). 
These values agree with line widths of other carbon-chain molecules \citep{2017ApJ...844...68T,2018ApJ...866..150T}.
In addition, these line widths are consistent with typical line widths of molecular lines in hot cores.
 
All of the three $^{13}$C isotopologues of HC$_{3}$N have been detected with S/N ratios above 10 in G12.89+0.49 and G16.86--2.16, and S/N ratios between 6 and 9 in G28.28--0.36 (Figure \ref{fig:HC3N}).
Their line widths are consistent with the main isotopologue of the same transition \citep[$J=10-9$;][]{2018ApJ...866..150T} within the errors.

Five $K-$ladder lines ($K=0-4$) of CH$_{3}$CN have been detected from G12.89+0.49 and G16.86--2.16, and four lines, missing the $J_{K}=5_{4}-4_{4}$ line, were detected from G28.28--0.36, as shown in the upper panels of Figure \ref{fig:CH3CN}.
The line widths of the $K=0$ line are in good agreement with those of carbon-chain species (CCH, HC$_{3}$N, and HC$_{5}$N), while the higher $K-$value lines show wider spectral features.
These results suggest that the lines with higher upper energy levels come from inner hot regions.

\floattable
\rotate
\begin{deluxetable}{lcccccccccccccccc}
\tabletypesize{\scriptsize}
\tablenum{2}
\tablecaption{Spectral line parameters \label{tab:line}}
\tablewidth{0pt}
\tablehead{
\colhead{} & \colhead{} & \colhead{} & \multicolumn{4}{c}{G12.89+0.49} & & \multicolumn{4}{c}{G16.86--2.16} & & \multicolumn{4}{c}{G28.28--0.36} \\
		\cline{4-7} \cline{9-12} \cline{14-17} 
		\colhead{Species} & \colhead{Frequency\tablenotemark{a}} & \colhead{$E_{\rm {up}}/k$} & \colhead{$T_{\rm {mb}}$} & \colhead{$\Delta v$\tablenotemark{b}} & \colhead{$\int T_{\rm {mb}}dv$} & \colhead{$V_{\rm {LSR}}$} & \colhead{} & \colhead{$T_{\rm {mb}}$} & \colhead{$\Delta v$\tablenotemark{b}} & \colhead{$\int T_{\rm {mb}}dv$} & \colhead{$V_{\rm {LSR}}$} & \colhead{} & \colhead{$T_{\rm {mb}}$} & \colhead{$\Delta v$\tablenotemark{b}} & \colhead{$\int T_{\rm {mb}}dv$} & \colhead{$V_{\rm {LSR}}$} \\
		\colhead{Transition} & \colhead{(GHz)} & \colhead{(K)} & \colhead{(K)} & \colhead{(\,km\,s$^{-1}$)} & \colhead{(K \,km\,s$^{-1}$)} & \colhead{(\,km\,s$^{-1}$)} &\colhead{} & \colhead{(K)} & \colhead{(\,km\,s$^{-1}$)} & \colhead{(K \,km\,s$^{-1}$)} & \colhead{(\,km\,s$^{-1}$)} & \colhead{} & \colhead{(K)} & \colhead{(\,km\,s$^{-1}$)} & \colhead{(K \,km\,s$^{-1}$)} &\colhead{(\,km\,s$^{-1}$)}
}
\startdata
		 CCH &  &  &  &  &  &  & &  &  &  &  & &  &  &  &  \\
		 $J=\frac{3}{2}-\frac{1}{2}$, &  &  &  &  &  &  & &  &  &  &  & &  &  &  &  \\
		 $F=1-1$ & 87.284156 & 4.2 &  0.341 (15) & 3.06 (16) & 1.15 (14) & 33.7 (1) & & 0.36 (3) & 3.1 (3) & 1.2 (2) & 17.3 (1) & & 0.359 (17) & 2.12 (12) & 0.86 (11) & 49.4 (1) \\ %CDMS
		 $F=2-1$ & 87.316925 & 4.2 &  1.548 (13) & 3.66 (4) & 6.2 (6) & 33.1 (1) & & 2.69 (3) & 2.73 (4) & 8.1 (8) & 17.5 (1) & & 1.272 (15) & 2.91 (4) & 4.1 (4) & 49.6 (1) \\
		 $F=1-0$ & 87.328624 & 4.2 &  1.042 (14) & 3.39 (5) & 3.9 (4) & 33.0 (1) & & 1.30 (3) & 3.38 (10) & 4.8 (5) & 17.4 (1) & & 0.852 (16) & 2.53 (6) & 2.4 (2) & 49.6 (1) \\
		 $J=\frac{1}{2}-\frac{1}{2}$, &  &  &  &  &  &  & &  &  &  &  & &  &  &  &  \\
		 $F=1-1$ & 87.402004 & 4.2 &  1.035 (14) & 3.36 (5) & 3.8 (4) & 33.5 (1) & & 1.28 (3) & 3.32 (10) & 4.7 (5) & 17.1 (1) & & 0.893 (16) & 2.52 (5) & 2.5 (3) & 49.2 (1) \\
		 $F=0-1$ & 87.407165 & 4.2 &  0.556 (15) & 3.04 (9) & 1.9 (2) & 32.8 (1) & & 0.65 (3) & 3.07 (19) & 2.2 (3) & 17.2 (1) & & 0.551 (17) & 2.15 (8) & 1.35 (15) & 49.3 (1)  \\
		 $F=1-0$ & 87.446512 & 4.2 &  0.336 (15) & 2.99 (16) & 1.11 (13) & 33.8 (1) & & 0.35 (3) & 3.1 (3) & 1.2 (2) & 17.3 (1) & & 0.335 (17) & 2.20 (14) & 0.83 (11) & 49.5 (1) \\
		 H$^{13}$CCCN &  &  &  &  &  &  & &  &  &  &  & &  &  &  &  \\
		  $J=10-9$ & 88.166832 & 23.3 & 0.103 (6) & 3.7(3) & 0.41 (6) & 32.4 (1) & & 0.092 (7) & 3.1 (3) & 0.31 (5) & 17.3 (1) & & 0.052 (9) & 2.4 (5) & 0.14 (4) & 49.1 (1) \\ %CDMS
		 HC$^{13}$CCN &  &  &  &  &  &  & &  &  &  &  & &  &  &  &  \\
		 $J=10-9$ & 90.593059 & 23.9 &  0.111 (5) & 3.9 (2) & 0.46 (6) & 33.3 (1) & & 0.100 (6) & 3.0 (2) & 0.33 (5) & 17.0 (1) & & 0.076 (9) & 1.6 (2) & 0.15 (3) & 49.0 (1) \\
		 HCC$^{13}$CN &  &  &  &  &  &  & &  &  &  &  & &  &  &  &  \\
		 $J=10-9$ & 90.601777 & 23.9 & 0.103 (5) & 3.7 (2) & 0.42 (5) & 32.3 (1) & & 0.096 (6) & 3.2 (3) & 0.33 (5) & 17.5 (1) & & 0.071 (8) & 2.3 (3) & 0.18 (4) & 49.5 (1) \\
		 CH$_{3}$CN &  &  &  &  &  &  & &  &  &  &  & &  &  &  &  \\ %JPL
		 $J_{K}=5_{4}-4_{4}$ & 91.9587260 & 127.5 & 0.116 (6) & 8.2 (5) & 1.01 (13) & 31.0 (1) & & 0.050 (8) & 9.5 (1.8) & 0.51 (14) & 17.3 (2) & & ... & ... & ... & ... \\
		 $J_{K}=5_{3}-4_{3}$ & 91.9711304 & 77.5 & 0.353 (7) & 6.23 (14) & 2.4 (2) & 34.1 (1) & & 0.188 (10) & 6.5 (4) & 1.3 (2) & 17.9 (1) & & 0.030 (6) & 4.5 (1.0) & 0.14 (5) & 50.3 (4) \\
		 $J_{K}=5_{2}-4_{2}$ & 91.9799943 & 41.8 & 0.410 (7) & 5.52 (11) & 2.4 (3) & 32.7 (1) & & 0.254 (11) & 5.3 (3) & 1.5 (2) & 17.4 (1) & & 0.075 (7) & 2.8 (3) & 0.23 (4) & 49.0 (2) \\
		 $J_{K}=5_{1}-4_{1}$ & 91.9853141 & 20.4 & 0.614 (7) & 5.88 (9) & 3.9 (4) &32.5 (2) & & 0.493 (12) & 4.83 (15) & 2.6 (3) & 17.2 (2) & & 0.219 (8) & 2.35 (10) & 0.58 (7) & 49.6 (3) \\
		 $J_{K}=5_{0}-4_{0}$ & 91.9870876 & 13.2 & 0.651 (9) & 3.68 (6) & 2.6 (3) & 32.8 (2) & & 0.587 (13) & 3.73 (11) & 2.4 (3) & 17.4 (2) & & 0.264 (8) & 2.10 (8) & 0.63 (7) & 49.0 (3) \\		 
\enddata
\tablecomments{Numbers in parentheses are the standard deviation, expressed in units of the last significant digits. The standard deviation for $T_{\rm {mb}}$ and $\Delta v$ is derived from the Gaussian fit, and we derived the standard deviation for $\int T_{\rm {mb}}dv$ taking the standard deviation of $T_{\rm {mb}}$ and $\Delta v$ and 10\% absolute calibration errors due to the chopper-wheel calibration into consideration.}
\tablenotetext{a}{Rest frequency and excitation energy are taken from the Cologne Database for Molecular Spectroscopy \citep[CDMS;][]{2005JMoSt.742..215M} for the lines of CCH and $^{13}$C isotopologues of HC$_{3}$N, and the Jet Propulsion Laboratory (JPL) catalog \citep{1998JQSRT..60..883P} for the CH$_{3}$CN lines.}
\tablenotetext{b}{Line widths are calculated using the following formula: $\Delta v = \sqrt{\Delta v_{\rm {gau}}^2-\Delta v_{\rm {inst}}^2}$, where $\Delta v_{\rm {gau}}$ and $\Delta v_{\rm {inst}}$ are the line widths obtained by the Gaussian fit and the instrumental velocity resolution (0.8 km s$^{-1}$), respectively.}
\end{deluxetable}

\subsection{Analyses} \label{sec:ana}

We describe analytical methods and present results in this subsection.
The derived column densities and excitation temperatures are summarized in Table \ref{tab:column}.
These derived column densities are the averaged values with a beam size of 18\arcsec, in the same manner as previous studies \citep{2017ApJ...844...68T, 2018ApJ...866..150T}.  

\subsubsection{CCH} \label{sec:anaCCH}

We derived excitation temperatures and column densities of CCH using the RADEX code \citep{2007A&A...468..627V}.
We assumed that the gas kinetic temperatures ($T_{\rm {gas}}$) are equal to the excitation temperatures of CH$_{3}$CCH, $T_{\rm {ex}}$(CH$_{3}$CCH), derived by the rotational diagram method using the $K-$ladder lines of the $J=5-4$ and $6-5$ transitions \citep{2018ApJ...866..150T}.
The $T_{\rm {ex}}$(CH$_{3}$CCH) values were derived to be $33^{+20}_{-9}$ K, $29_{-8}^{+15}$ K, and $23_{-6}^{+9}$ K in G12.89+0.49, G16.86--2.16, and G28.28--0.36, respectively.
The line widths of CCH are comparable to those of CH$_{3}$CCH, which supports the hypothesis that these species trace similar temperature regions and our assumption that $T_{\rm {gas}}$ can be set at $T_{\rm {ex}}$(CH$_{3}$CCH).

The chemical model calculations \citep{2019ApJ...881...57T} indicate that carbon-chain species are formed in the lukewarm  ($T>25$ K) region, where the CH$_{4}$ molecules are desorbed from dust grains, as in WCCC regions. 
In such temperature ranges, the lowest H$_{2}$ density was estimated to be $\sim10^{5}$ cm$^{-3}$ based on hot-core models by \citet{2004A&A...414..409N}.
We then assumed that the H$_{2}$ density is $1 \times 10^{5}$ cm$^{-3}$ at first.

With the above conditions ($n$(H$_{2}$)$ =1 \times 10^{5}$ cm$^{-3}$ and $T_{\rm {gas}}=33$ K, 29 K, and 23 K in G12.89+0.49, G16.86--2.16, and G28.28--0.36, respectively), the derived excitation temperatures of CCH in G12.89+0.49 and G16.86--2.16 are higher than the assumed $T_{\rm {gas}}$ values (case a in Table \ref{tab:column}).
In G28.28--0.36, the column density and excitation temperature of CCH are ($2.8 \pm 1.1$)$\times 10^{14}$ cm$^{-2}$ and $16 \pm 3$ K.

Next, we reduced the H$_{2}$ density to $5 \times 10^{4}$ cm$^{-3}$ and derived the column densities and excitation temperatures of CCH (case b in Table \ref{tab:column}).
The column densities and excitation temperatures of CCH are derived to be ($4.1 \pm 1.4$)$\times 10^{14}$ cm$^{-2}$ and $13 \pm 2$ K in G12.89+0.49, and ($5.1 \pm 1.0$)$\times 10^{14}$ cm$^{-2}$ and $11 \pm 2$ K in G16.86--2.16, respectively.
The column density of CCH is derived to be ($3.1 \pm 1.3$)$\times 10^{14}$ cm$^{-3}$ in G28.28--0.36.
The column densities of CCH do not change and are in agreement for cases a and b within their errors.

Finally, we tried $T_{\rm {gas}}$ values at the lower limits of $T_{\rm {ex}}$(CH$_{3}$CCH) - 24 K in G12.89+0.49, 21 K in G16.86--2.16, and 17 K in G28.28--0.36 - and an H$_{2}$ density of $1 \times 10^{5}$ cm$^{-3}$ (case c in Table \ref{tab:column}).
The derived column densities of CCH do not change significantly among all of the sets of assumed physical parameters.
The results are consistent within their errors in all of the sources.

In the density range between $5\times10^{4}$ cm$^{-3}$ and $10^{6}$ cm$^{-3}$, the derived column densities of CCH increase by only a factor of two at $n$(H$_{2}$)$=10^{6}$ cm$^{-3}$.
Thus, the assumed physical parameters do not change the column densities of CCH significantly, and our following discussions are not affected.

\subsubsection{$^{13}$C isotopologues of HC$_{3}$N} \label{sec:anaHC3N}

\citet{2018ApJ...866..150T} derived the column densities and rotational temperatures of the main isotopologue of HC$_{3}$N, including the $J=10-9$ line, toward the same MYSOs using the rotational diagram method under LTE conditions by following \citet{1999ApJ...517..209G}. 
Here, we assume that the spatial distributions of the $^{13}$C isotopologues are the same as the main isotopologue and apply the LTE assumption for simplicity.
We used the following formulae \citep{2016ApJ...817..147T}:
\begin{equation} \label{tau}
\tau = - {\mathrm {ln}} \left[1- \frac{T_{\rm {mb}} }{J(T_{\rm {ex}}) - J(T_{\rm {bg}})} \right]
\end{equation}
where
\begin{equation} \label{tem}
J(T) = \frac{h\nu}{k}\Bigl\{\exp\Bigl(\frac{h\nu}{kT}\Bigr) -1\Bigr\} ^{-1},
\end{equation}  
and
\begin{equation} \label{col}
N = \tau \frac{3h\Delta v}{8\pi ^3}\sqrt{\frac{\pi}{4\mathrm {ln}2}}Q\frac{1}{\mu ^2}\frac{1}{J_{\rm {lower}}+1}\exp\Bigl(\frac{E_{\rm {lower}}}{kT_{\rm {ex}}}\Bigr)\Bigl\{1-\exp\Bigl(-\frac{h\nu }{kT_{\rm {ex}}}\Bigr)\Bigr\} ^{-1}.
\end{equation} 
In Equation (\ref{tau}), $\tau$ denotes the optical depth, and $T_{\rm {mb}}$ represents the peak intensities summarized in Table \ref{tab:line}.
The excitation temperature and the cosmic microwave background temperature ($\simeq 2.73$ K) are indicated as $T_{\rm{ex}}$ and $T_{\rm {bg}}$, respectively.
We assumed that the excitation temperatures of the $^{13}$C isotopologues of HC$_{3}$N are equal to the rotational temperatures of the main isotopologue \citep{2018ApJ...866..150T}; 24 K, 20 K, and 13.4 K in G12.89+0.49, G16.86--2.16, and G28.28--0.36, respectively.
Since we have considered the latest result on $T_{\rm{ex}}$ for HC$_{3}$N \citep{2018ApJ...866..150T} as the most plausible, we used their results here.
$J$($T$) in Equation (\ref{tem}) is the effective temperature equivalent to that in the Rayleigh-Jeans law.
The symbols of $h$, $k$, and $\nu$ denote the Planck constant, Boltzmann constant, and rest frequency, respectively.
In Equation (\ref{col}), {\it N} is the column density,  $\Delta v$ is the line width (Table \ref{tab:line}), $Q$ is the rotational partition function, $\mu$ is the permanent electric dipole moment, and $E_{\rm {lower}}$ is the energy of the lower rotational energy level. 
We assume that the electric dipole moments of the $^{13}$C isotopologues of HC$_{3}$N are equal to that of the main isotopologue \citep[3.73172 D;][]{1985JChPh..82.1702D}.

The derived column densities are summarized in Table \ref{tab:column}.
In G12.89+0.49 and G16.86--2.16, the column densities of the three $^{13}$C isotopologues are consistent with each other in each source.
On the other hand, HCC$^{13}$CN is slightly more abundant than the others in G28.28--0.36, which agrees with the previous results \citep{2016ApJ...830..106T}.

In these observations, we observed only one rotational line for each $^{13}$C isotopomer, and derived the column densities by fixing the excitation temperatures.
Hence, the absolute values of column densities may contain extra uncertainties due to the assumed $T_{\rm {ex}}$ value.
However, these uncertainties do not affect our discussion, because we focus on the $^{13}$C isotopic fractionation of HC$_{3}$N, or relative differences in abundance among the $^{13}$C isotopologues, which are expected to trace the same conditions.
Relative abundances are not affected by the assumed excitation temperature or absolute column densities, when we compare them using the same transition.

The derived column densities in G28.28--0.36 reported here and those of \citet{2016ApJ...830..106T} differ.
This is caused by the different assumed excitation temperatures.
We believe that the column densities derived here are more reliable, because \citet{2016ApJ...830..106T} assumed that the excitation temperature is 100 K, a typical hot core temperature.
As mentioned before, the different values of column density do not affect our discussion.
Our final conclusion using $^{13}$C isotopic fractionation (Section \ref{sec:dis3}) is consistent with that of \citet{2016ApJ...830..106T}.

\subsubsection{CH$_{3}$CN} \label{sec:anaCH3CN}

We derived the rotational temperatures and column densities of CH$_{3}$CN in the three sources from a rotational diagram analysis, using the following formula \citep{1999ApJ...517..209G};
\begin{equation} \label{rd}
{\rm {ln}} \frac{3k \int T_{\mathrm {mb}}dv}{8\pi ^3 \nu S \mu ^2} = {\rm {ln}} \frac{N}{Q(T_{\rm {rot}})} - \frac{E_{\rm {up}}}{kT_{\rm {rot}}},
\end{equation}
where $S$ is the line strength, $E_{\rm {up}}$ is the upper energy level, and $Q(T_{\rm {rot}})$ is the partition function for the rotational temperature $T_{\rm {rot}}$.
The electric dipole moment of CH$_{3}$CN is 3.92197 D \citep{1998JQSRT..60..883P}.
The $\int T_{\mathrm {mb}}dv$ integral represents the integrated intensity values summarized in Table \ref{tab:line}.

The rotational diagrams for each source are shown in the bottom panels of Figure \ref{fig:CH3CN}.
In the case of G12.89+0.49 and G16.86--2.16, we excluded the $J_{K}=5_{4}-4_{4}$ line ($E_{\rm {up}}/k=127.5$ K) from the fitting, because the fitting results are significantly bad if we include this point.
We conjecture that contributions of inner hot regions for this line are larger than those for other lines. 
In fact, the line widths of the $J_{K}=5_{4}-4_{4}$ line are significantly larger than those of other CH$_{3}$CN lines (Table \ref{tab:line}).
The derived column densities and rotational temperatures are $1.2_{-0.4}^{+1.2} \times 10^{14}$ cm$^{-2}$ and $119_{-36}^{+85}$ K in G12.89+0.49, $4.5_{-0.9}^{+1.8} \times 10^{13}$ cm$^{-2}$ and $68_{-15}^{+26}$ K in G16.86--2.16, and $5.1_{-0.8}^{+1.1} \times 10^{12}$ cm$^{-2}$ and $36_{-8}^{+11}$ K in G28.28--0.36, respectively (see Table \ref{tab:column}).

We do not know sizes of the emission regions and so could not apply a beam dilution correction for each line. 
The analyses without the beam dilution correction may overestimate the rotational temperature and underestimate the column density.

\begin{deluxetable*}{lcccccccc}
\tablenum{3}
\tablecaption{Column density and excitation temperature in three MYSOs\label{tab:column}}
\tablewidth{0pt}
\tablehead{
                 \colhead{Species} & \multicolumn{2}{c}{G12.89+0.49} & & \multicolumn{2}{c}{G16.86--2.16} & & \multicolumn{2}{c}{G28.28--0.36} \\
		       \cline{2-3} \cline{5-6} \cline{8-9}
		    \colhead{}  & \colhead{$N$ (cm$^{-2}$)} & \colhead{$T_{\rm {ex}}$ (K)} & \colhead{} & \colhead{$N$ (cm$^{-2}$)} & \colhead{$T_{\rm {ex}}$ (K)} & \colhead{} & \colhead{$N$ (cm$^{-2}$)} & \colhead{$T_{\rm {ex}}$ (K)} 
}
\startdata
		CCH\tablenotemark{a} & ($4.2 \pm 1.4$)$\times 10^{14}$ & $40 \pm 15$ & & ($4.9 \pm 1.1$)$\times 10^{14}$ & $26 \pm 6$ & & ($2.8 \pm 1.1$)$\times 10^{14}$ & $16 \pm 3$ \\
		CCH\tablenotemark{b} & ($4.1 \pm 1.4$)$\times 10^{14}$ & $13 \pm 2$ & & ($5.1 \pm 1.0$)$\times 10^{14}$ & $11 \pm 2$ & & ($3.1 \pm 1.3$)$\times 10^{14}$ & $9 \pm 1$ \\
		CCH\tablenotemark{c} & ($3.9 \pm 1.3$)$\times 10^{14}$ & $17 \pm 3$ & & ($4.8 \pm 0.9$)$\times 10^{14}$ & $14 \pm 2$ & & ($2.9 \pm 1.2$)$\times 10^{14}$ & $11 \pm 1$ \\
		H$^{13}$CCCN & ($1.8 \pm 0.2$)$\times 10^{12}$ & 24\tablenotemark{d} & & ($1.4 \pm 0.2$)$\times 10^{12}$ & 20\tablenotemark{d} & & ($7.6 \pm 2.1$)$\times 10^{11}$ & 13.4\tablenotemark{d} \\
		HC$^{13}$CCN & ($1.9 \pm 0.2$)$\times 10^{12}$ & 24\tablenotemark{d} & & ($1.4 \pm 0.2$)$\times 10^{12}$ & 20\tablenotemark{d} & & ($7.9 \pm 1.6$)$\times 10^{11}$ & 13.4\tablenotemark{d} \\
		HCC$^{13}$CN & ($1.8 \pm 0.2$)$\times 10^{12}$ & 24\tablenotemark{d} & & ($1.4 \pm 0.2$)$\times 10^{12}$ & 20\tablenotemark{d} & & ($9.97 \pm 1.9$)$\times 10^{11}$ & 13.4\tablenotemark{d} \\
		CH$_{3}$CN & $1.2_{-0.4}^{+1.2} \times 10^{14}$ & $119_{-36}^{+85}$ & & $4.5_{-0.9}^{+1.8} \times 10^{13}$ & $68_{-15}^{+26}$ & & $5.1_{-0.8}^{+1.1} \times 10^{12}$ & $36_{-8}^{+11}$ \\
\enddata
		\tablecomments{The errors are the standard deviation.}
		\tablenotetext{a}{The assumed gas kinetic temperatures are fixed at the excitation temperatures of CH$_{3}$CCH derived by \citet{2018ApJ...866..150T}: 33 K, 29 K, and 23 K for G12.89+0.49, G16.86--2.16, and G28.28--0.36, respectively. The assumed H$_{2}$ densities are fixed at $1 \times 10^{5}$ cm$^{-3}$.}
		\tablenotetext{b}{The assumed gas kinetic temperatures are fixed at the excitation temperatures of CH$_{3}$CCH derived by \citet{2018ApJ...866..150T}. The assumed H$_{2}$ densities are fixed at $5\times 10^{4}$ cm$^{-3}$.}
		\tablenotetext{c}{The assumed gas kinetic temperatures are fixed at the lower limits of excitation temperatures of CH$_{3}$CCH derived by \citet{2018ApJ...866..150T}; 24 K, 21 K, and 17 K for G12.89+0.49, G16.86--2.16, and G28.28--0.36, respectively. The assumed H$_{2}$ densities are fixed at $1 \times 10^{5}$ cm$^{-3}$.}
		\tablenotetext{d}{The excitation temperatures are fixed at the main species derived by \citet{2018ApJ...866..150T}.}
\end{deluxetable*}

\section{Discussion} \label{sec:dis}

\subsection{Comparison of the CCH/HC$_{5}$N ratio between MYSOs and WCCC sources} \label{sec:dis1}

\begin{deluxetable*}{ccccccc}
\tablenum{4}
\tablecaption{Comparison of the CCH/HC$_{5}$N ratio around young protostars\label{tab:ratio}}
\tablewidth{0pt}
\tablehead{
		\colhead{Source Type} & \multicolumn{3}{c}{MYSOs} & & & \colhead{Low-mass WCCC source} \\
		\cline{2-4} \cline{6-7}
		\colhead{Source}   & \colhead{G12.89+0.49\tablenotemark{a}} & \colhead{G16.86--2.16\tablenotemark{a}} & \colhead{G28.28--0.36\tablenotemark{a}} & \colhead{} & \colhead{} & \colhead{L1527\tablenotemark{b}} 
}
\startdata
		CCH/HC$_{5}$N & $17^{+8}_{-7}$ & $17^{+6}_{-5}$ & $14^{+6}_{-7}$ & & & $625^{+3041}_{-339}$ \\ 
\enddata
              \tablecomments{The upper and lower errors were calculated as [$N$(CCH)$\pm\delta$$N$(CCH)]/[$N$(HC$_{5}$N)$\mp\delta$$N$(HC$_{5}$N)], where $\delta$$N$(species) represents the standard deviation of the column densities.}
              \tablenotetext{a}{The column densities of HC$_{5}$N were taken from \citet{2017ApJ...844...68T}; $2.39^{+0.15}_{-0.17} \times 10^{13}$ cm$^{-2}$, $2.78^{+0.16}_{-0.2} \times 10^{13}$ cm$^{-2}$, and $2.05^{+0.2}_{-0.05} \times 10^{13}$ cm$^{-2}$ in G12.89+0.49, G16.86--2.16, and G28.28--0.36, respectively.}
		\tablenotetext{b}{The column densities of HC$_{5}$N and CCH were taken from \citet{2019PASJ...71S..18Y}.} 
\end{deluxetable*}

As mentioned in Section \ref{sec:intro}, the CCH/HC$_{5}$N ratio can probably be used as a temperature probe for regions where carbon-chain molecules exist.
In this subsection, we will compare the CCH/HC$_{5}$N ratio among MYSOs and a low-mass WCCC source. 

We summarize the CCH/HC$_{5}$N ratio around both the three high-mass protostars and one low-mass protostar (L1527) in Table \ref{tab:ratio}.
We estimated the errors for the most severe cases to confirm that the CCH/HC$_{5}$N ratios in MYSOs are clearly different from that in L1527.

The column densities of HC$_{5}$N were previously derived to be $2.39^{+0.15}_{-0.17} \times 10^{13}$ cm$^{-2}$, $2.78^{+0.16}_{-0.2} \times 10^{13}$ cm$^{-2}$, and $2.05^{+0.2}_{-0.05} \times 10^{13}$ cm$^{-2}$ in G12.89+0.49, G16.86--2.16, and G28.28--0.36, respectively \citep{2017ApJ...844...68T}.
These column densities were derived by the rotational diagram method using ten lines with upper state energies of 7.0 -- 99.7 K.
Using the column densities of CCH summarized in Table \ref{tab:column}, the CCH/HC$_{5}$N ratios are calculated to be $17^{+8}_{-7}$, $17^{+6}_{-5}$, and $14^{+6}_{-7}$ in each MYSO.
These high-mass values were obtained using the column densities of CCH in case ``c'' in Table \ref{tab:column}. 
The errors cover all of the three cases.
These CCH/HC$_{5}$N ratios are the averaged values over the Nobeyama beam size of 18\arcsec.

For L1527, the low-mass WCCC source, \citet{2019PASJ...71S..18Y} derived the column density and excitation temperature of HC$_{5}$N to be ($2.4 \pm 1.8$)$\times10^{12}$ cm$^{-2}$ and $22 \pm 7$ K, respectively.
\citet{2019PASJ...71S..18Y} derived the CCH column density in L1527 to be (1.2--2.2)$\times10^{15}$ cm$^{-2}$ and ($1.5 \pm 0.3$)$\times10^{15}$ cm$^{-2}$ by non-LTE and LTE methods, respectively.
Thus, the CCH/HC$_{5}$N ratio in L1527 is $625^{+3041}_{-339}$.

We clearly recognize a difference in the CCH/HC$_{5}$N ratio between MYSOs and the L1527 low-mass WCCC source (see Table \ref{tab:ratio}).
Specifically, the CCH/HC$_{5}$N ratios in MYSOs are lower than that in the low-mass WCCC source by one order of magnitude at least, suggesting that the carbon-chain chemistry around MYSOs is different from that in low-mass counterparts.

\subsection{Temperature dependence of the CCH/HC$_{5}$N ratio} \label{sec:dis2}

\begin{figure}[!th]
\figurenum{4}
 \begin{center}
  \includegraphics[bb = 0 20 334 686, scale=0.7]{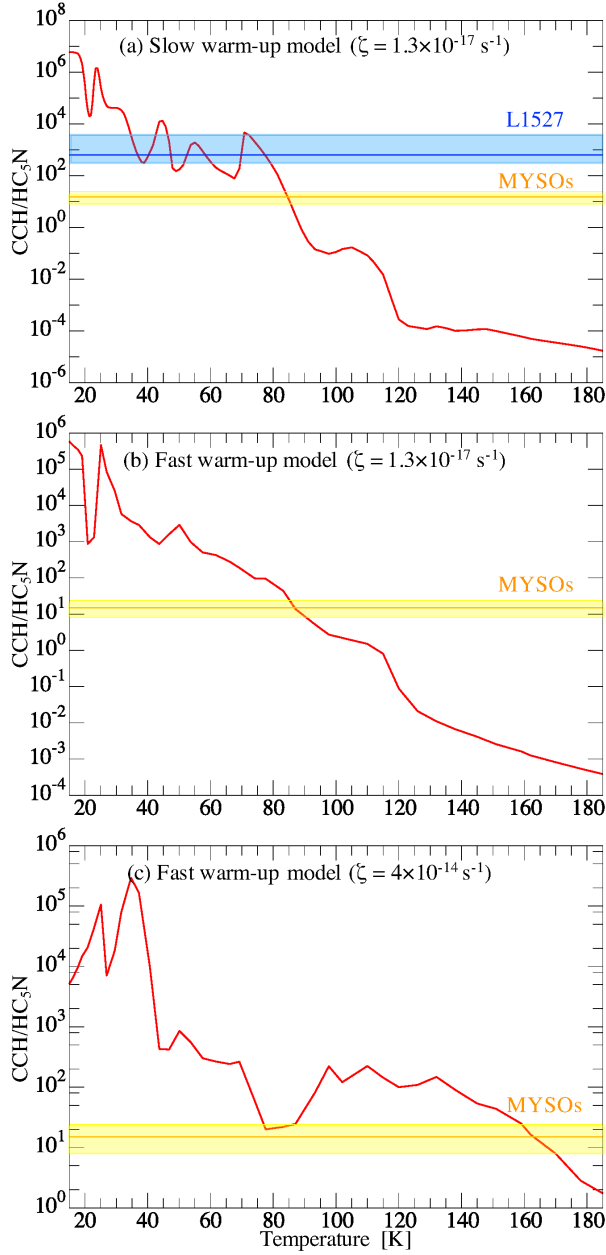}
 \end{center}
\caption{Comparisons of the CCH/HC$_{5}$N ratio between the observed values and the model calculations \citep{2019ApJ...881...57T}. The red curves show the modeled CCH/HC$_{5}$N ratio in (a) the Slow warm-up period ($1 \times 10^{6}$ yr) model with $\zeta=1.3\times10^{-17}$ s$^{-1}$, (b) the Fast warm-up period ($5 \times 10^{4}$ yr) model with $\zeta=1.3\times10^{-17}$ s$^{-1}$ and (c) the Fast warm-up period ($5 \times 10^{4}$ yr) model with $\zeta=4.0\times10^{-14}$ s$^{-1}$. The orange line indicates the observed value in G28.28--0.36 and the yellow range covers the observed values in the three MYSOs including the standard deviation. The blue line in panel (a) indicates the observed value in L1527 and the blue range covers the observed error in L1527.\label{fig:model1}}
\end{figure}

We now compare the observed CCH/HC$_{5}$N ratios with the modeled values.
\citet{2019ApJ...881...57T} ran time-dependent chemical simulations of hot-core models with a warm-up period using the state-of-the-art three-phase chemical code Nautilus \citep{2016MNRAS.459.3756R}.
In this section, we first use the results with a slow warm-up period ($1\times10^{6}$ yr) between 15 and 185 K, because our primary goal is to investigate the temperature dependence of the CCH/HC$_{5}$N ratio in detail.
Later we will use a faster warm-up period for comparison.

Panel (a) of Figure \ref{fig:model1} shows comparisons between the observed CCH/HC$_{5}$N ratios in MYSOs (orange horizontal line) and L1527 (blue horizontal line) along with the modeled value as a function of temperature.
The modeled CCH/HC$_{5}$N ratio generally decreases as the temperature increases.
The observed CCH/HC$_{5}$N values in the MYSOs agree best with the modeled value at $T\simeq 85$ K. 
In the case of L1527, there are some regions where the observed value is reproduced by the model calculation, but the most likely point is at $T\simeq 35$ K, taking the distributions of carbon-chain species in L1527 \citep{2010ApJ...722.1633S} and the profile of the gas temperature \citep[e.g.,][]{2018A&A...615A..83V} into consideration.
This temperature almost agrees with the carbon-chain spatial distribution in L1527 \citep{2010ApJ...722.1633S} and matches the WCCC model \citep{2008ApJ...681.1385H}.

Panel (b) of Figure \ref{fig:model1} shows the result of a fast warm-up period ($5 \times 10^{4}$ yr) with the same cosmic ray ionization rate ($1.3\times10^{-17}$ s$^{-1}$) as Panel (a).
The observed CCH/HC$_{5}$N ratio agrees with the model at $T\simeq 87$ K, which is consistent with the temperature suggested from Panel (a) of $\approx 85$ K.

To summarize, carbon-chain species around MYSOs exist in higher temperature regions than those in low-mass WCCC sources.
This suggestion is also supported by the facts that the CCH column densities in the MYSOs are lower than those in the low-mass WCCC sources, while the HC$_{5}$N column densities in the MYSOs are higher than those in the low-mass counterparts by more than one order of magnitude.
These results may indicate that HC$_{5}$N exists in higher temperature regions where CCH is deficient around the MYSOs, as discussed in detail later.

Distributions of carbon-chain species could be extended to cold and/or lukewarm envelopes too \citep[e.g.,][]{2019ECS.....3.2659B}.
In that case, the observed CCH/HC$_{5}$N ratios should be considered as the upper limits because of the mixing with low-temperature components.
Such contributions from the low-temperature components are expected to be larger in MYSOs than those in low-mass protostars.
This means that the temperature where carbon-chain species exist around MYSOs estimated from the above comparison is the lower limit, and our conclusion that carbon-chain molecules around MYSOs exist in higher temperature regions compared to those around low-mass WCCC sources does not change.
In fact, \citet{2019ApJ...881...57T} suggested that the observed lower limit of the HC$_{5}$N abundance and the observed HC$_{5}$N/CH$_{3}$OH ratio in G28.28--0.36 can be reproduced when the temperature reaches the desorption temperature of HC$_{5}$N ($T \approx 115$ K).
Thus, HC$_{5}$N around MYSOs could exist in hot-core regions with temperatures above 100 K.

Comparisons of ratios between molecular column densities are more reliable than those of fractional abundances with respect to H$_{2}$, mainly because of large uncertainties in deriving H$_{2}$ column densities.
Hence, we conclude that HC$_{5}$N around MYSOs exists in higher temperature regions than in the low-mass WCCC case.
In such temperature regimes, CCH seems to be destroyed efficiently by reactions with O ($T \leq 90$ K) and H$_{2}$ ($T \geq 90$ K).
On the other hand, the gas-phase HC$_{5}$N molecules are removed mainly by reactions with HCO$^{+}$ and by adsorption onto dust grains.
The gas-phase HCO$^{+}$ abundance is much lower than those of O and H$_{2}$ by a few orders of magnitude.
Therefore, the destruction rate of HC$_{5}$N is slower than that of CCH in such high temperature conditions.
In fact, the CCH column densities in MYSOs are lower than those in low-mass WCCC sources, while the HC$_{5}$N column densities in MYSOs are higher than those in low-mass WCCC sources.
This implies that CCH is possibly destroyed more efficiently in high-mass star forming regions in larger scales because of the higher temperature.
A similar conclusion that CCH is efficiently destroyed around massive stars has also been reached in previous studies \citep[e.g.,][]{2015ApJ...808..114J}.
Moreover, the possibility of more efficient destruction of carbon-chain species by atomic oxygen in high-mass star-forming regions compared to that in low-mass counterparts was suggested by \citet{2018ApJ...854..133T}.

Panel (c) of Figure \ref{fig:model1} shows the model result with a high cosmic ray ionization rate of $4 \times 10^{-14}$ s$^{-1}$ and a fast warm-up period ($5 \times 10^{4}$ yr) taken from \citet{2019ApJ...881...57T}.
We present only the fast warm-up model because the observed HC$_{5}$N abundance could not be reproduced in the slow warm-up model and the high cosmic-ray ionization rate \citep[see][]{2019ApJ...881...57T}.
Such a high cosmic ray ionization rate could be possible in protostellar systems \citep{2016A&A...590A...8P}, and abundances of a few carbon-chain species in the OMC-2 FIR 4 young intermediate-mass protoclusters can be reproduced  with this high cosmic ray ionization rate \citep{2017A&A...605A..57F,2018ApJ...859..136F}.
In this model, the observed CCH/HC$_{5}$N ratios in MYSOs can be explained at temperatures of $\sim 80$ K and $\sim 160$ K. 
Since the observed CCH/HC$_{5}$N ratios are possibly the upper limits as discussed before, temperatures above 160 K are not omitted, while the observed ratios cannot be reproduced by the model at temperatures of $\sim 90-160$ K.
Again, the suggested temperatures where carbon-chain species exist ($\simeq 80$ K and $\geq 160$ K) are higher than those in the low-mass WCCC source.

\subsection{$^{13}$C Isotopic Fractionation of HC$_{3}$N in MYSOs} \label{sec:dis3}

During the observations presented in this paper, we have detected the three $^{13}$C isotopologues of HC$_{3}$N in the three MYSOs.
We summarize the derived $^{12}$C/$^{13}$C ratios of HC$_{3}$N in the three MYSOs in Table \ref{tab:HC3Nratio}.
The $^{12}$C/$^{13}$C ratio depends on the distance from the Galactic Center, hereafter $D_{\rm {GC}}$ \citep[e.g.,][]{2005ApJ...634.1126M}. 
In Table \ref{tab:HC3Nratio}, we listed the predicted $^{12}$C/$^{13}$C ratios derived by the following formula \citep{2019ApJ...877..154Y}:
\begin{equation}
^{12}\mathrm{C}/^{13}\mathrm{C} = (5.08 \pm 1.10) \times D_{\rm {GC}} + (11.86 \pm 6.60).
\end{equation}
The $D_{\rm {GC}}$ values are estimated at 5.2 kpc, 6.4 kpc, and 5.5 kpc for G12.89+0.49, G16.86--2.16, and G28.28--0.36, respectively, applying the law of cosines and assuming that the distance between the Galactic Center and the Sun is 8 kpc \citep{2003ApJ...597L.121E}.
The observed $^{12}$C/$^{13}$C ratios in the three MYSOs almost agree with the predicted values within their errors.
This means that the dilution of the $^{13}$C species does not occur for HC$_{3}$N, which was also found in nearby low-mass starless cores \citep{2017ApJ...846...46T, 2019ApJ...884..167T}.
The $^{12}$C/$^{13}$C ratios will change if the excitation temperatures of the $^{13}$C isotopologues are different from those of the $^{12}$C species.
Thus, we do not discuss the matter in further detail, because we derived the column densities of the $^{13}$C isotopologues with fixed excitation temperatures (Section \ref{sec:anaHC3N}).

\begin{deluxetable}{cccc}
\tablenum{5}
\tablecaption{$^{12}$C/$^{13}$C ratio of HC$_{3}$N in the three MYSOs\label{tab:HC3Nratio}}
\tablewidth{0pt}
\tablehead{
		  \colhead{}  & \colhead{G12.89+0.49} & \colhead{G16.86--2.16} & \colhead{G28.28--0.36}
}
\startdata
		H$^{13}$CCCN & $25^{+4}_{-3}$ & $32^{+5}_{-4}$ & $31^{+6}_{-10}$ \\
		HC$^{13}$CCN & $23^{+3}_{-2}$ & $31 \pm 4$ & $30^{+2}_{-9}$ \\
		HCC$^{13}$CN & $25^{+4}_{-3}$ & $30 \pm 4$ & $24^{+1}_{-7}$ \\
		Average & 24 & 31 & 28 \\
		Prediction\tablenotemark{a} & 26--51 & 31 -- 58 & 27 -- 52 \\
\enddata
             \tablecomments{The column densities of the main isotopologue were derived to be $4.4 \times 10^{13}$ cm$^{-2}$, $4.3 \times 10^{13}$ cm$^{-2}$, and $2.0 \times 10^{13}$ cm$^{-2}$ in G12.89+0.49, G16.86--2.16, and G28.28--0.36, respectively \citep{2018ApJ...866..150T}. The errors are the standard deviation.}
		\tablenotetext{a}{Calculated using the formula derived by \citet{2019ApJ...877..154Y}.}
\end{deluxetable}

\begin{deluxetable}{cl}
\tablenum{6}
\tablecaption{The $^{13}$C isotopic fractionation of HC$_{3}$N in the three MYSOs\label{tab:13Cfrac}}
\tablewidth{0pt}
\tablehead{
		  \colhead{Source}  & \colhead{[H$^{13}$CCCN] : [HC$^{13}$CCN] : [HCC$^{13}$CN]} 
}
\startdata
G12.89+0.49 & 0.92 ($\pm 0.12$) : 1.00 : 0.90 ($\pm 0.11$) \\
G16.86--2.16 & 0.96 ($\pm 0.14$) : 1.00 : 1.00 ($\pm 0.14$) \\
G28.28--0.36 & 1.0 ($\pm 0.3$) : 1.0 : 1.3 ($\pm 0.2$) \\
\enddata
             \tablecomments{The errors are the standard deviation.}
\end{deluxetable}

We derived the column density ratios of [H$^{13}$CCCN] : [HC$^{13}$CCN] : [HCC$^{13}$CN], namely the $^{13}$C isotopic fractionation, and the results are summarized in Table \ref{tab:13Cfrac}. 
\citet{2016ApJ...830..106T} determined these ratios to be 1.0 ($\pm 0.2$) : 1.00 : 1.47 ($\pm 0.17$) ($1 \sigma$) in G28.28--0.36 with higher velocity resolution (0.5 km s$^{-1}$) spectra.
The ratios derived in this paper in G28.28--0.36 show a similar tendency; the abundances of H$^{13}$CCCN and HC$^{13}$CCN are comparable to each other, and the abundance of HCC$^{13}$CN is higher than the others.
On the other hand, in G12.89+0.49 and G16.86--2.16, we could not recognize any differences in abundance among the $^{13}$C isotopologues.

Based on the $^{13}$C isotopic fractionation patterns, we can constrain the main formation pathways of carbon-chain species \citep[e.g.,][]{2016ApJ...817..147T, 2018MNRAS.474.5068B}. 
Regarding HC$_{3}$N, three fractionation patterns and corresponding formation pathways were proposed \citep{2016ApJ...830..106T, 2017ApJ...846...46T}:
\begin{enumerate}
\item If the abundance ratios are [H$^{13}$CCCN] : [HC$^{13}$CCN] : [HCC$^{13}$CN] = $1 : 1 : x$ ($x$ is an arbitrary value), the reaction ``C$_{2}$H$_{2}$ + CN" is dominant.
\item If the abundance ratios are [H$^{13}$CCCN] : [HC$^{13}$CCN] : [HCC$^{13}$CN] = $y : 1 : z$ ($y$ and $z$ are arbitrary values), the reaction ``CCH + HNC" is dominant.
\item If the abundance ratios are [H$^{13}$CCCN] : [HC$^{13}$CCN] : [HCC$^{13}$CN] $\approx 1 : 1 : 1$, the electron recombination reaction of HC$_{3}$NH$^{+}$ is dominant.
\end{enumerate}
The third pathway is a hypothesis, because \citet{2017ApJ...846...46T} assumed that there are competitive formation pathways of HC$_{3}$NH$^{+}$, which have not been confirmed by any observations.
According to the above classification, the main formation pathway of HC$_{3}$N is the electron recombination reaction of HC$_{3}$NH$^{+}$ in G12.89+0.49 and G16.86--2.16.
In G28.28--0.36, the reaction between C$_{2}$H$_{2}$ and CN is dominant, which agrees with the conclusion by \citet{2016ApJ...830..106T}.
We will discuss a possible explanation for the difference in the main formation mechanism of HC$_{3}$N between G28.28--0.36 and the others in the next subsection.

\subsection{Constraints of ages of MYSOs based on the CCH/HC$_{5}$N ratio and the $^{13}$C isotopic fractionation of HC$_{3}$N} \label{sec:dis4}

\begin{figure}[!th]
\figurenum{5}
 \begin{center}
  \includegraphics[bb= 0 30 346 706, scale=0.7]{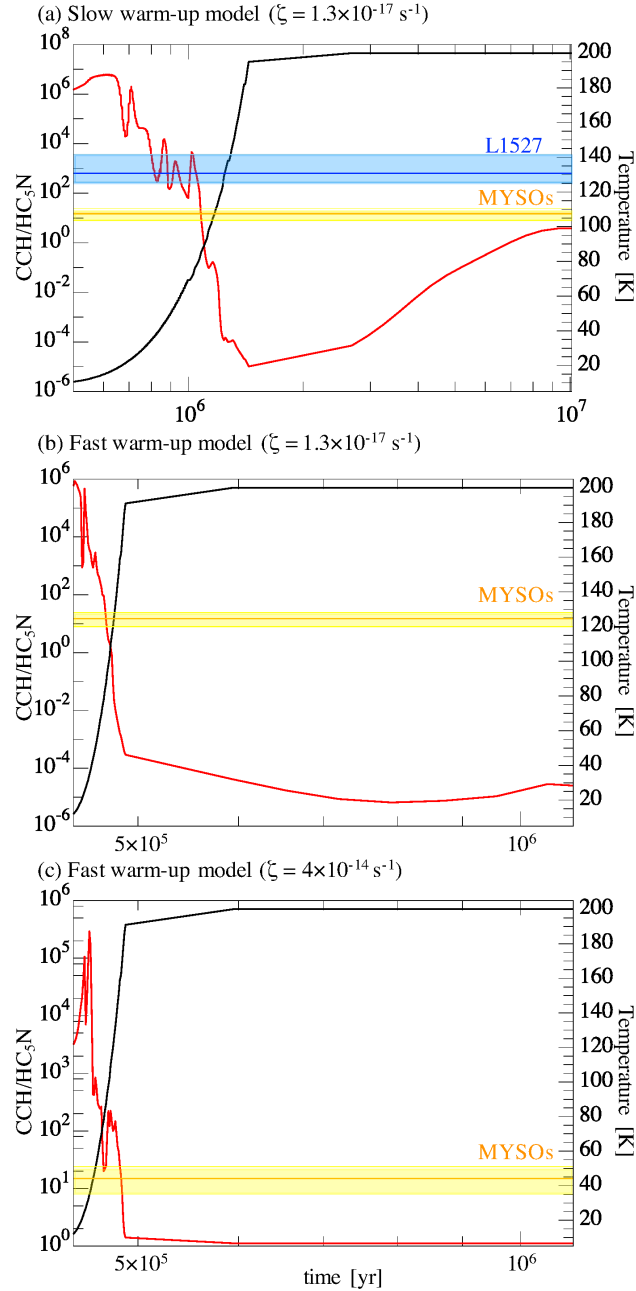}
 \end{center}
\caption{Comparisons of the CCH/HC$_{5}$N ratio between the observed values and the model calculations \citep{2019ApJ...881...57T}. The red and black curves show the modeled CCH/HC$_{5}$N ratio and temperature, respectively, in (a) the Slow warm-up period model with $\zeta=1.3\times10^{-17}$ s$^{-1}$, (b) the Fast warm-up period ($5 \times 10^{4}$ yr) model with $\zeta=1.3\times10^{-17}$ s$^{-1}$, and (c) the Fast warm-up period model with $\zeta=4.0\times10^{-14}$ s$^{-1}$. The orange line indicates the observed value in G28.28--0.36 and the yellow range covers the observed values in the three MYSOs including the standard deviation. The blue line in panel (a) indicates the observed value in L1527 and the blue range covers the error in L1527.\label{fig:model2}}
\end{figure}

\begin{figure}[!th]
\figurenum{6}
 \begin{center}
  \includegraphics[bb=0 30 370 590, scale=0.57]{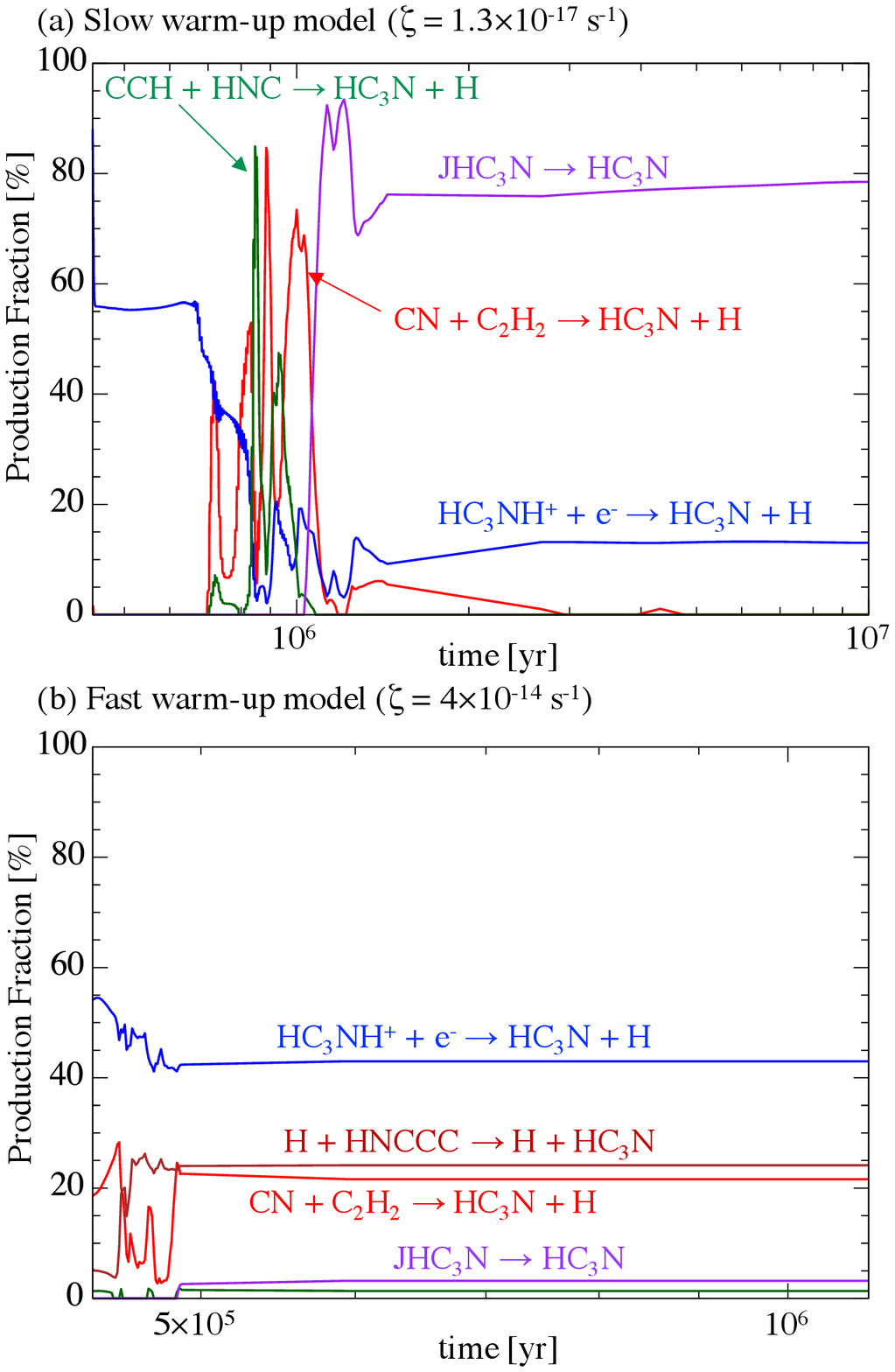}
 \end{center}
\caption{Production fraction of formation pathways of HC$_{3}$N during the warm-up period \citep{2019ApJ...881...57T}. JHC$_{3}$N denotes molecules on the dust surface. \label{fig:model_HC3N}}
\end{figure}

We constrain the age and possible carbon-chain chemistry in each MYSO based on the CCH/HC$_{5}$N ratio (Section \ref{sec:dis2}) and the $^{13}$C isotopic fractionation of HC$_{3}$N (Section \ref{sec:dis3}).
Figure \ref{fig:model2} shows the time dependence of the CCH/HC$_{5}$N ratio and temperature during the warm-up period \citep{2019ApJ...881...57T}.
The data are the same as in Figure \ref{fig:model1}.
Figure \ref{fig:model_HC3N} shows the time dependence of contributions for each formation pathway of HC$_{3}$N.
The basic formation and destruction mechanisms of carbon-chain species do not change between the different warm-up periods with the same cosmic ray ionization rate ($1.3 \times 10^{-17}$ s$^{-1}$) \citep{2019ApJ...881...57T}.
Since the formation and destruction mechanisms depend on the temperature, the age value will change but our discussions need not change.
We can investigate the time-dependence of the formation pathways of HC$_{3}$N in detail using the model with the slow warm-up period, and so presented this model here.
The results of the fast warm-up period with a cosmic ray ionization rate of $1.3 \times 10^{-17}$ s$^{-1}$ were not presented in Figure \ref{fig:model_HC3N}, because we were not able to find any new results.

As discussed in Section \ref{sec:dis2}, the observed CCH/HC$_{5}$N ratios in MYSOs agree with the model at a temperature of $\sim 85$ K.
This temperature corresponds to $1.08 \times 10^{6}$ yr\footnote{This age does not mean a physical age of a MYSO, because we derive the age from the chemical network simulations and the age will change by the assumed warm-up timescales.}, as shown in panel (a) of Figure \ref{fig:model2}. 
Just before this time ($t < 1.07 \times 10^{6}$ yr), the main formation pathway of HC$_{3}$N is the reaction between C$_{2}$H$_{2}$ and CN (panel (a) of Figure \ref{fig:model_HC3N}).
This is consistent with the reaction proposed based on the $^{13}$C isotopic fractionation in G28.28--0.36.

The electron recombination reaction of HC$_{3}$NH$^{+}$ is dominant in the model with the high cosmic ray ionization rate during the whole warm-up period (panel (b) of Figure \ref{fig:model_HC3N}).
Thus, the $^{13}$C isotopic fractionation of HC$_{3}$N in G12.89+0.49 and G16.86--2.16 can be explained by this high cosmic ray ionization rate model.
These results suggest that the cosmic ray ionization rates in G12.89+0.49 and G16.86--2.16 may be higher than that in G28.28--0.36 or physical structures in G12.89+0.49 and G16.86--2.16 allow cosmic rays to penetrate into regions where carbon-chain species are formed.
The observed CCH/HC$_{5}$N ratios in MYSOs agree with the age of (4.7--4.86)$\times 10^{5}$ yr as indicated in panel (c) of Figure \ref{fig:model2}.  

In summary, the observed CCH/HC$_{5}$N ratios around MYSOs can be reproduced in the chemical network simulations at a given temperature and time.
Taking the $^{13}$C isotopic fractionation of HC$_{3}$N into consideration, G12.89+0.49 and G16.86--2.16 may prefer the model with the high cosmic ray ionization rate ($\zeta = 4 \times 10^{-14}$ s$^{-1}$), while G28.28--0.36 agrees with the model with the standard cosmic ray ionization rate  ($\zeta = 1.3 \times 10^{-17}$ s$^{-1}$). 
These results imply that carbon-chain chemistry in G12.89+0.49 and G16.86--2.16 may resemble that in the OMC-2 FIR 4 young intermediate-mass protoclusters \citep{2017A&A...605A..57F,2018ApJ...859..136F}, but that in G28.28--0.36 is different from the other MYSOs and OMC-2 FIR 4.
The clear $^{13}$C isotopic fractionation of HC$_{3}$N in G28.28--0.36 supports its bottom-up formation during the warm-up stage, while carbon-chain species, at least HC$_{5}$N, exist in higher temperature regions than those in low-mass WCCC sources.
Different chemical pathways are present because of the possibly different physical conditions, as suggested by our chemical code.
Future observations with interferometers such as ALMA will clarify such a new type of carbon-chain chemistry around MYSOs.

\section{Conclusions} \label{sec:con}

We have carried out observations of the rotational lines of CCH ($N=1-0$), CH$_{3}$CN ($J=5-4$), and three $^{13}$C isotopologues of HC$_{3}$N ($J=10-9$) toward the three MYSOs, G12.89+0.49, G16.86--2.16, and G28.28--0.36, with the Nobeyama 45-m telescope.
The observational results and main conclusions are as follows:
\begin{enumerate}
\item  We determined the CCH/HC$_{5}$N ratios, which are considered as a temperature probe where carbon-chain species exist in the three MYSOs.
The CCH/HC$_{5}$N ratios are derived to be $\sim 15$ in the MYSOs.
These CCH/HC$_{5}$N ratios in the MYSOs are lower than those in low-mass WCCC sources by more than one order of magnitude.

\item We compare these observational values to the chemical network simulations with a warm-up period.
The observed CCH/HC$_{5}$N ratios in the MYSOs are reproduced when the temperature reaches $\sim 85$ K, while that in L1527 agrees with the model at a temperature of $\sim 35$ K.
In the model calculation with the high cosmic ray ionization rate ($4 \times 10^{-14}$ s$^{-1}$), the observed CCH/HC$_{5}$N ratios in the MYSOs can be reproduced at temperatures of $\sim 80$ K and $\sim 160$ K.
Hence, HC$_{5}$N detected around the MYSOs exists preferentially in higher temperature regions than that in low-mass WCCC sources.
In such temperature regimes ($T \geq 70$ K), CCH, which is a reactive species, is efficiently destroyed by reactions with O and/or H$_{2}$.

\item We determined the $^{13}$C isotopic fractionation of HC$_{3}$N in the three MYSOs.
All of the three $^{13}$C isotopologues of HC$_{3}$N show similar column densities in G12.89+0.49 and G16.86--2.16, while HCC$^{13}$CN is more abundant than the others in G28.28--0.36.
Based on the results, the electron recombination reaction of HC$_{3}$NH$^{+}$ is proposed as the main formation pathway of HC$_{3}$N in G12.89+0.49 and G16.86--2.16.
This is consistent with the model with the high cosmic ray ionization rate.
On the other hand, in G28.28--0.36, the main formation pathway of HC$_{3}$N is the reaction between C$_{2}$H$_{2}$ and CN.
This reaction is dominant in the model with the standard cosmic ray ionization rate ($1.3 \times 10^{-17}$ s$^{-1}$).
\end{enumerate}

Based on the CCH/HC$_{5}$N ratio and the $^{13}$C isotopic fractionation, the carbon-chain chemistry in the G28.28--0.36 MYSO seems to be different from that in the other two MYSOs, OMC-2 FIR 4 \citep{2017A&A...605A..57F,2018ApJ...859..136F}, and low-mass WCCC sources.
In this source, cyanopolyynes are likely formed by the bottom-up mechanism during the lukewarm regions ($25 < T < 50$ K) and exist in higher temperature regions ($T \gtrsim 85$ K) than in low-mass WCCC sources.

\acknowledgments
We would like to express our special thanks to the staff of the Nobeyama Radio Observatory.
The Nobeyama Radio Observatory is a branch of the National Astronomical Observatory of Japan, National Institutes of Natural Sciences.
K.T. would like to thank the University of Virginia for providing the funds for her postdoctoral fellowship in the Virginia Initiative on Cosmic Origins (VICO) research program.
This work was supported by JSPS KAKENHI Grant Number JP20K14523.
E.H. thanks the National Science Foundation for support through grant AST-1906489.
ZYL is supported in part by NSF AST-1910106 and NASA 80NSSC20K0533.
JCT acknowledges ERC grant MSTAR and VR grant 2017-04522.

%% To help institutions obtain information on the effectiveness of their 
%% telescopes the AAS Journals has created a group of keywords for telescope 
%% facilities.
%
%% Following the acknowledgments section, use the following syntax and the
%% \facility{} or \facilities{} macros to list the keywords of facilities used 
%% in the research for the paper.  Each keyword is check against the master 
%% list during copy editing.  Individual instruments can be provided in 
%% parentheses, after the keyword, but they are not verified.

\vspace{5mm}

\facilities{Nobeyama 45-m radio telescope}

%% Similar to \facility{}, there is the optional \software command to allow 
%% authors a place to specify which programs were used during the creation of 
%% the manuscript. Authors should list each code and include either a
%% citation or url to the code inside ()s when available.

\software{NEWSTAR, Nautilus \citep{2016MNRAS.459.3756R}}

%% Appendix material should be preceded with a single \appendix command.
%% There should be a \section command for each appendix. Mark appendix
%% subsections with the same markup you use in the main body of the paper.

%% Each Appendix (indicated with \section) will be lettered A, B, C, etc.
%% The equation counter will reset when it encounters the \appendix
%% command and will number appendix equations (A1), (A2), etc. The
%% Figure and Table counter will not reset.

%\appendix

%% This command is needed to show the entire author+affiliation list when
%% the collaboration and author truncation commands are used.  It has to
%% go at the end of the manuscript.
%\allauthors

%% Include this line if you are using the \added, \replaced, \deleted
%% commands to see a summary list of all changes at the end of the article.
%\listofchanges

\end{document}